\documentclass[preprint,showpacs,preprintnumbers,amsmath,amssymb]{revtex4}
\usepackage{txfonts}
\usepackage{tipa}

\usepackage{graphicx}
\usepackage{dcolumn}

\begin{document}

\title{Voltage-Controlled Berry Phases in Two Coupled Quantum Dots}

\author{Huan Wang}
\email{wanghuan2626@sjtu.edu.cn}
\author{Ka-Di Zhu}
\email{  zhukadi@sjtu.edu.cn }
\affiliation{ Department of Physics,
Shanghai Jiao Tong University, Shanghai 200240, P.R.China }

\date{\today}

\begin{abstract}
The voltage-controlled Berry phases in two vertically coupled
InGaAs/GaAs quantum dots are investigated theoretically. It is found
that Berry phases can be changed dramatically from 0 to 2$\pi$ (or
2$\pi$ to 0) only simply by turning the external voltage. Under
realistic conditions, as the tunneling is varied from $0.8eV$ to
$0.9eV$ via a bias voltage, the Berry phases are altered obviously,
which can be detected in an interference experiment. The scheme is
expected to be useful in constructing quantum computation based on
geometric phases in an asymmetrical double quantum dot controlled by
voltage.
\end{abstract}

\maketitle Recently with the advent of quantum information and
communication\cite{Nilsen}, the phase of a wavefunction plays an
important role in numerous quantum information protocols. The state
vector of a quantum system can rotate as it undergoes a cyclic
evolution in state space, such that it returns to its initial
physical state, its wavefunction can acquire a geometric phase
factor in addition to the familiar dynamic phase\cite{Segao,
Anandan}. If the cyclic change of the system is adiabatic, this
additional factor is known as Berry's phase\cite{Berry}. Since it
has potential applications in the implementation of quantum
computation by geometric means\cite{Jones,Falci}, which is less
susceptible to noise from the environment. Therefore the study on
Berry phase is becoming more and more important. Fuentes-Guridi et
al.\cite{Fuentes} calculated the Berry phase of a particle in a
magnetic field considering the quantum nature of the field. Yi et
al.\cite{Yi} studied the Berry phase in a composite system and
showed how the Berry phases depend on the coupling between the two
subsystems. San-Jose et al.\cite{San} have described the effect of
geometric phases induced by either classical or quantum electric
fields acting on single electron spins in quantum dots. Yuan and
Zhu\cite{Yuan} have shown that the Berry phases of two coupled
quantum dots depend on the environmental temperatures. Most
recently, observations of Berry phases in solid state materials are
reported\cite{Yuanbo,Vartiainen,Leek}. Leek et al.\cite{Leek}
demonstrated the controlled Berry phase in a superconducting qubit
which manipulates the qubit geometrically using microwave radiation
and observes the phase in an interference experiment. In this
letter, we theoretically present a scheme where the Berry phases can
be controlled by a bias voltage in a double quantum dot (QD). The
Berry phases can change dramatically from 0 to $2\pi$ (or from
$2\pi$ to 0) only simply by applying the external voltage. This
scheme is expected to be useful in constructing quantum computation
based on geometric phase in an asymmetrical double quantum dot
controlled by voltage.

\begin{figure}[h]
\centering
\includegraphics[width=15cm]{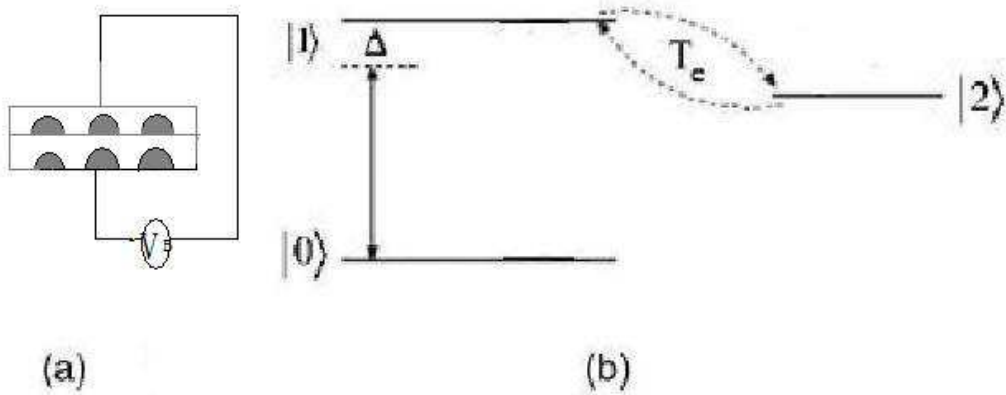}
\caption{(a)Schematic of the setup. An optical pulse transmits the
left QD. $V_B$ is a bias voltage. (b)Schematic band structure and
level configuration of a double QD system for the
electromagnetically induced transparency (EIT) operation. A pulse
laser excites one electron from the valence band that can tunnel to
other dot,and $|0\rangle$ is the system without excitations,
$|1\rangle$ a pair os electron and hole bound in the first dot, and
$|2\rangle$ one hole in the first dot with an electron in he second
dot. } \label{fig:sect1:learning}
\end{figure}

 A vertically coupled InGaAs/GaAs asymmetrical quantum dot molecule
consists of two layers of dots (the upper one and the lower one)
with different band structures coupled by tunnelling is shown in
Fig.1(a). Samples are arrays of GaInAs dots in a matrix of GaAs
which are vertically stacked, vertically aligned, and electrically
coupled in the growth direction. Dots in two different layers show a
strong tendency to align vertically. The coupling is mainly
determined by the separation distance of two layers. In this quantum
dot system,  the lower QD is slightly small, which the energy
separation is bigger than the upper one. From the
Ref.~\cite{Krenner}, we know that for QD separation $d>9$ nm the
tunnelling coupling between the two dots is weak and the QD system
can be discussed in terms of a simplified single-particle
picture\cite{Chun, Bester}.  One can excite one electron from the
valence to the conduction band in the lower dot which can in turn
tunnel to the upper dot by applying electromagnetical field. Figure
1 (a) gives a schematic of the system. The tunnel barrier in an
asymmetric quantum dot molecule can be controlled by applying a bias
voltage between the $n^+$-contact and the Schottky gate. Figure 1(b)
depicts an energy-level diagram of an asymmetric quantum dot
molecule. The ground state $|0\rangle$ is the system without
excitations, and the direct exciton state $|1\rangle$ is a pair of
electron and hole bound in the lower dot, and the indirect exciton
state $|2\rangle$ is one hole in the lower dot with an electron in
the upper dot. Using this configuration the Hamiltonian of the
system reads as follows($\hbar=1$)\cite{Villas}

\begin{equation}
\begin{split}
H_{1}=&\sum_{j=0}^{2} \varepsilon_j | j \rangle \langle j |
+\omega_c
a_c^+a_c+g_c(|1\rangle\langle0|a_c+|0\rangle\langle1|a_c^+)\\
&+T_c(|1\rangle\langle2|+|2\rangle\langle1|),
\end{split}\end{equation}
where $\varepsilon_j$ is the energy of state $|j\rangle$, $T_c$ is
the electron tunnelling matrix element between two dots, $a^{+}_{c}$
and $a_{c}$ are, respectively, the creation and annihilation
operators of the quantized field with frequency $\omega_c$. $g_{c}$
is the coupling constant of the quantized field and the direct
exciton (the state of $|0\rangle$ and $|1\rangle$). The Hamiltonian
$H_1$ is rather simple, but it is not complete. Since the double
quantum dots is embedded in the macroscopic crystal, the single
electron is unavoidably scattered by phonons while tunnelling
between two dots. Considering the coupling between electron and
phonons, the Hamiltonian can be written as

\begin{equation}
H=H_1 + \sum_k \omega_k b_k^
+b_k+\\
\frac{1}{2}(|1\rangle\langle1|-|2\rangle\langle2|)\sum_k
g_k(b_k^++b_k),
\end{equation}
where $b_k^+$ $(b_k)$ and $\omega_k$ are the creation (annihilation)
operator and energy for $k$th phonon mode, respectively, $g_k$ is
the coupling constant determined by the crystal material and the
geometry of the coupling quantum dots. The electron-phonon
interaction in Hamiltonian (2) contains only the diagonal elements,
because the role of off-diagonal ones is suppressed at low
temperatures. Applying a canonical transformation with the
generator\cite{Zhuo, Du}

\begin{equation}
S=(|1\rangle\langle1|-|2\rangle\langle2|)\sum_k\frac{g_k}{2\omega_k}(b_k^+-b_k).
\end{equation}

The transformed Hamiltonian is given by
\begin{equation}
H'=e^sHe^{-s}=H_0'+H_I',
\end{equation}
where
\begin{equation}
\begin{split}
H_0'=&\varepsilon_0|0\rangle\langle0|+(\varepsilon_1-\Delta)|1\rangle\langle1|+
(\varepsilon_2-\Delta)|2\rangle\langle2|\\
&+\sum_k\omega_k b_k^+b_k+\omega_c a_c^+a_c
\end{split}
\end{equation}
\begin{equation}
H_I'=g_c(|1\rangle\langle0|X^+a_c+|0\rangle\langle1|Xa_c^+)+T_c(|2\rangle\langle1|X^2+
|1\rangle\langle2|X^{2+})
\end{equation}
\begin{equation}
\Delta=\sum_k\frac{g_k^2}{4\omega_k},X=exp[-\sum_k\frac{g_k}{2\omega_k}(b_k^+-b_k)].
\end{equation}
Here we assume that the relaxing time of the environment (phonon
fields) is so short that the excitons do not have time to exchange
the energy and information with the environment before the
environment returns to its equilibrium state. The excitons interact
weakly with the environment so that the equilibrium thermal
properties of the environment are preserved. Therefore it is
reasonable to replace the operator $X$ with its expectation value
over the phonons number states which are determined by  a thermal
average and write the Hamiltonian as\cite{Mahan,Chen,Zhu1}

\begin{equation}
\begin{split}
H'=
 &\varepsilon0|0\rangle\langle0|+(\varepsilon1-\Delta)|1\rangle\langle1|+
(\varepsilon_2-\Delta)|2\rangle\langle2|\\
 &+\sum_k\omega_k b_k^+
b_k+\omega_c a_c^+ a_c\\
 &+g_c
e^{-\lambda(N_{ph}+1/2)}(|1\rangle\langle0|a_c+|0\rangle\langle1|a_c^+)\\
&+
T_ce^{-2\lambda(N_{ph}+1/2)}(|2\rangle\langle1|+|1\rangle\langle2|)
\end{split}
\end{equation}
where $\lambda=\sum_k(g_{k}/2\omega_{k})^2$ is the Huang-Rhys
factor. Here for the sake of simplicity we only perform the analysis
for the simplest case in which only the longitudinal-optical (LO)
phonon is considered,i.e., all the phonons have the same frequency
\cite{Yuan,Chen,Zhu1}. We anticipate that this is sufficient to
illustrate the main physics in the more complicated case of the
acoustic phonons\cite{Du}. In such a case $\omega_k=\omega_0$ is
irrelevant to the wavevector $k$ of phonon, and the phonon
populations can be written as
$N_{ph}=\frac{1}{e\frac{\omega_0}{k_{B}T}-1}$\cite{Mahan,Chen}where
$k_B$ is Boltzmann constant and $T$ is temperature of the
environment. After using the operator $\Lambda=e^{i\sum_k\omega_0
b_k^+ b_k}$ to transform to a frame rotating at the frequency
$\omega_0$, we can cancel the term $\sum_k\omega_0 b_k^+ b_k$ in
Eq.(8) and get its eigenstates (l=1,2,3)

\begin{equation}
|\psi\rangle^l=C_0^l|0,n+1\rangle+C_1^l|1,n\rangle+C_2^l|2,n\rangle,
\end{equation}
where $n$ is the photon number of the quantized field,

$C_0^l=\frac{1}{\sqrt{1+a_l^2+b_l^2}}$,
$C_1^l=\frac{a_l}{\sqrt{1+a_l^2+b_l^2}}$,  and
$C_2^l=\frac{b_l}{\sqrt{1+a_l^2+b_l^2}}$.  Here

\begin{equation}
a_l=\frac{\eta_l-e_1}{e_4},
\end{equation}

\begin{equation}
b_l=\frac{e_5}{e_4}\frac{\eta_l-e_1}{\eta_l-e_3},
\end{equation}
with
$e_1=\varepsilon_0+(n+1)\omega_c$,$e_2=\varepsilon_1-\Delta+n\omega_c$,
$e_3=\varepsilon_2-\Delta+n\omega_c$,$e_4=g_ce^{-\lambda(N_{ph}+1/2)}\sqrt{n+1}$,
$e_5=T_c e^{-2\lambda(N_{ph}+1/2)}$. Also $\eta1$,$\eta2$ and
$\eta3$ are the roots of the equation

\begin{equation}
\begin{split}
&\eta^3-(e_1+e_2+e_3)\eta^2+(e_1 e_2+e_1 e_3+e_2
e_3-e_4^2-e_5^2)\eta\\
 &+ (e_1 e_5^2+e_4^2 e_3-e_1 e_3 e_3)=0.
\end{split}
\end{equation}

According to Ref.\cite{Fuentes} ,the phase shift operator
$U(\varphi)=e^{-\varphi a^+a}$ is introduced. Applied adiabatically
to the Hamiltonian Eq.(8), the phase shift operator alters the sates
of the field and gives rise to the following eigenstates:

\begin{equation}
|\psi\rangle_p^l=C_0^le^{-i(n+1)\varphi}|0,n+1\rangle+
C_1^le^{-in\varphi}|1,n\rangle+C_2^le^{-in\varphi}|2,n\rangle
\end{equation}
Changing $\varphi$ slowly from 0 to 2$\pi$ , the Berry phase is
calculated as $\gamma_l=i \int_0^{2\pi}
{_p}{^l}\langle\psi|\frac{\partial}{\partial\varphi}|\psi\rangle_p^l
d\varphi,$  it gives $\gamma_l=2\pi[n+(C_0^l)^{2}]$.

\begin{figure}[h] 
\includegraphics[width=15cm]{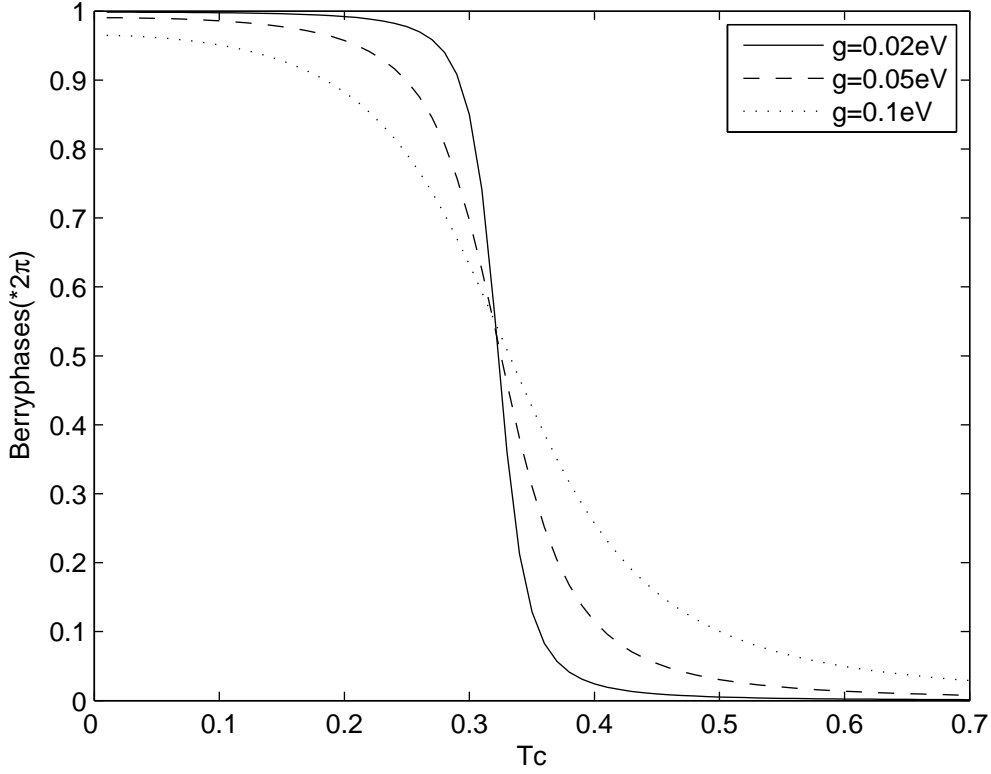}
\caption{The Berry phases of the system as the function of $T_c$
with different $\gamma$,($\gamma_1$ solid curve,$\gamma_2$ dashed
curve)Where $n=0$, $\varepsilon_0=0eV$, $\varepsilon_1=1.3eV$,
$\varepsilon_2=1.0eV$, $g=0.02eV$, $\Delta=0.1eV$
$\omegaup_c=0.3eV$, $\lambda=0.02eV $, $T=0$
}\label{fig:sect1:learning3}
\end{figure}

\begin{figure}[h]
\includegraphics[width=15cm]{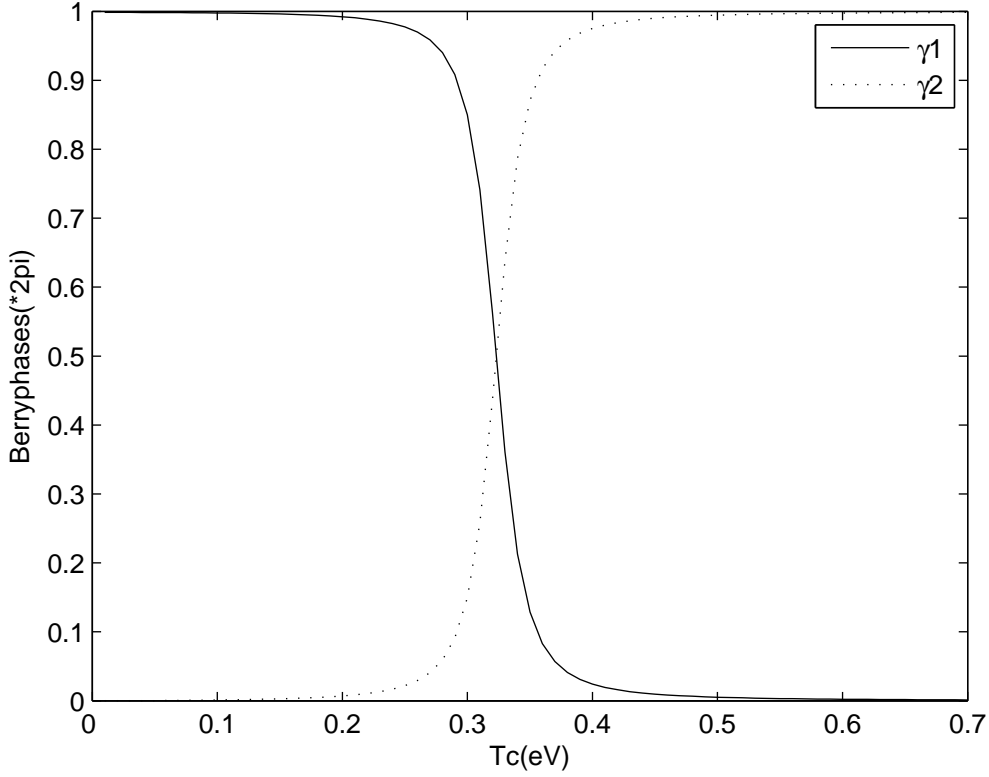}
\caption{The Berry phases the function of $T_c$ with different
$g_c$, Where $n=0$, $\varepsilon_0=0eV$, $\varepsilon_1=1.3eV$,
$\varepsilon_2=1.0eV$, $\Delta=0.1eV$ $\omegaup_c=0.3eV$,
$\Delta=0.02eV, $T=0$ $}\label{fig:sect1:learning}
\end{figure}

In calculation, $\varepsilon_0=0meV$,$\varepsilon_1=1.3meV$ and
$\varepsilon_2=1.0meV$ are be used\cite{Legramd,Findeis,Grandidier}.
Figure1 shows the variation Berry phases when the tunneling is
varied via gate voltage. It is obvious that the Berry phase is
different from zero even for the driving field in the vacuum state
($n=0$). As the parameters g chosen, the Berry phase is an
approximately 2m$\pi$(m is an integer)shift as the tunneling
$T_c<0.3eV$ or $T_c>0.4eV $. As the result of quantizing the driving
field, the term $2m\pi$ is appears in the Berry phase. The range of
$T_c$ where the Berry phase changes obviously from $2m\pi$ to
$2(m\pm1)\pi$ is related to different coupling constants g. As the
coupling constants increase, the changing of the Berry phase are
inclining to become slow. From the picture we can see that the Berry
phase is changing more sharp when g is 0.01eV than g is 0.1eV.
 Figure2 show two kinds of Berry phases. There are three Berry phases
in all, here we just give two typical kinks of them($\gamma_1,
\gamma_2$). The $\gamma_3$ does not give a new results. Figure 2
shows the two curves about $\gamma_1$ and $\gamma_2$ are symmetric.

For the illustration of the numerical results, we choose the
vertically coupled InGaAs/GaAs quantum dots as an example for
experiments. For such a double dot, we assume that $\varepsilon_0=0
eV$,$\varepsilon_1=1.3eV$ and $\varepsilon_2=1.0eV$\cite{Krenner1},
$\Delta=0.01eV$ and $\lambda=0.02$\cite{Kamada}. The LO-phonon
energy ($\omega_{0}$) of GaAs is 36 meV\cite{Mahan}. We apply a
quantized field with frequency $\omega_c=0.3eV$ which is just
resonant with the bonding state of the level $|1\rangle$ and the
level $|2\rangle$ via the tunnel coupling as shown in Fig.1(a).
Figure 2 shows the Berry phases as a function of tunnelling ($T_c$)
via a bias voltage for three coupling constants ($g_c$) at $T=50mK$.
It is obvious that the Berry phase is different from zero even for
the driving field in the vacuum state ($n=0$), which is in agreement
with the results obtained by Fuentes-Guridi et al.\cite{Fuentes}. By
simply tuning by the bias voltage, the Berry phases can be changed
dramatically from 2$\pi$ to 0 as the parameter $g_c$ is fixed. From
Fig.1 we can see that the Berry phase is an approximately 2$\pi$ for
the tunnelling $T_c<0.8eV$, but as the tunnelling continuously
increases to $T_c>0.9eV$ the Berry phase is suddenly down to zero.
As realized in experiments, we can fix the incident light through a
single mode fiber and continuously tuning the bias voltage, if the
Berry phase is suddenly changed, then we can detect this predicted
effect in an interference experiment. The range of $T_c$ where the
Berry phase changes obviously from $2m\pi$ to $2(m\pm1)\pi$ ($m$ is
an integer) is related to the different coupling constants ($g_c$).
As the coupling constants increase, the change of the Berry phase
becomes slow. From the figure it is evident that the Berry phases
are altered more sharply as $g_c= 0.02eV$ than $g_c=0.1eV$.
 Figure 3 shows two kinds of Berry phases ($\gamma_1,
\gamma_2$) as a function of tunnelling $T_c$. In general, there are
three Berry phases in all, here we just give two typical kinds of
them ($\gamma_1, \gamma_2$). The $\gamma_3$ does not give a new
result. Figure 3 shows the two curves about $\gamma_1$ and
$\gamma_2$ are symmetrical at $T_c\approx0.85eV$. It should be noted
here that the present scheme can also applied to the double quantum
dot system realized by Gustavsson et al.\cite{Gus}, which the
driving field is operated at microwave frequency. In such a case,
the tunnelling $T_c$ can be reached to $meV$ or even to $\mu eV$
through varying the gate voltage.

In conclusion, we have theoretically investigated the
voltage-controlled Berry phases in an asymmetry semiconductor double
quantum dots. It is found that Berry phases can be changed suddenly
from 0 to 2$\pi$ (or 2$\pi$ to 0) only simply by tuning the external
voltage. Under realistic experimental conditions, as the tunnelling
is varied from $0.8eV$ to $0.9eV$ via a bias voltage, the Berry
phase can be altered dramatically. The range of $T_c$ where the
Berry phase changes obviously from $2m\pi$ to $2(m\pm1)\pi$ is
related to different coupling constant ($g_c$). As the coupling
constant $g_c$ enhances, the change of the Berry phases become slow.
This scheme opens up the electrical controllability of the Berry
phases which is expected to be useful in constructing quantum
computer based on geometric phases in an asymmetric double quantum
dot controlled by voltage.

This work has been supported in part by National Natural Science
Foundation of China (No.10774101) and the National Ministry of
Education Program for Training Ph.D.

\end{document}